\documentclass[journal]{IEEEtran}
\usepackage[utf8]{inputenc}
\usepackage{graphicx}
\ifCLASSOPTIONcompsoc
\usepackage[caption=false, font=normalsize, labelfont=sf, textfont=sf]{subfig}
\else
\usepackage[caption=false, font=footnotesize]{subfig}
\fi
\usepackage{amssymb}
\usepackage{amsthm}
\usepackage{amsmath}
\usepackage{makecell}

\newtheorem{theorem}{Theorem}[section]
\newtheorem{proposition}[theorem]{Proposition}

\begin{document}

\title{Dynamic Population Estimation\\ Using Anonymized Mobility Data}
\author{
	Xiang Liu and Philo Pöllmann
	\thanks{The authors are with HERE Technologies, Invalidenstraße 116, Berlin 10115, Germany (email: xiang.liu@here.com).}
}

\maketitle

\begin{abstract}
Fine population distribution both in space and in time is crucial for epidemic management, disaster prevention, urban planning and more. Human mobility data have a great potential for mapping population distribution at a high level of spatiotemporal resolution. Power law models are the most popular ones for mapping mobility data to population. However, they fail to provide consistent estimations under different spatial and temporal resolutions, i.e. they have to be recalibrated whenever the spatial or temporal partitioning scheme changes. We propose a Bayesian model for dynamic population estimation using static census data and anonymized mobility data. Our model gives consistent population estimations under different spatial and temporal resolutions.
\end{abstract}


\section{Introduction}

Traditionally, a nationwide census is carried out every 5 to 10 years and each individual is enumerated at the residence, such as US~\cite{us:census} and Germany~\cite{germany:census:2011}. While a census is the most important data source of static population, it has a few limitations. Firstly, such a census needs to be planned well in advance and takes at least months to complete, which is expensive and time-consuming. Secondly, the census population usually reports all residents at a reference date, regardless whether they are physically there on that day. Thirdly, the census population is static in nature, which does not meet growing demands for fine spatiotemporal population data. For example, institutional monitoring of the effectiveness of lockdown measures during epidemic.

Several projects have focused on improving the spatial resolution of census data, namely Gridded Population of the World (GPW)~\cite{gpw}, Global Rural Urban Mapping Project (GRUMP)~\cite{grump}, Global Human Settlement Layer - Population (GHS-POP)~\cite{ghs-pop}, WorldPop~\cite{worldpop}, World Population Estimation~\cite{wpe,frye:settlement:2018} and LandScan~\cite{landscan}. These datasets provide gridded population at a global level and a spatial resolution from 100~m to 1~km. While some of these datasets interpolate population between two consecutive census years, none of them has estimated population dynamics within a day. A comprehensive review on gridded population datasets is provided in~\cite{popgrid,leyk:review:2019}.

A variety of socioeconomic and morphological variables have been found correlated to census population density and have been used for improving static population estimation, such as nightlight~\cite{sutton:ambient:2003,bharti:overlaying:2015}, land use~\cite{stevens:disaggregating:2015,bakillah:openstreetmap:2014,douglass:resolution:2015}, point of interests~\cite{bakillah:openstreetmap:2014,yao:building:2017}, building footprints and volumes~\cite{biljecki:netherlands:2016}, etc. Random forest is a popular model for static population estimation~\cite{stevens:disaggregating:2015,douglass:resolution:2015,patel:geotweet:2017,yao:building:2017}, which combines multiple socioeconomic and morphological layers easily. However, these variables are not able to reveal fine population dynamics due to their essentially static property. Recently human mobility data, such as mobile phone activity and social network activity, are used for both static~\cite{khodabandelou:metadata:2019,patel:geotweet:2017,yao:building:2017,botta:crowd:2015,steiger:twitter:2015,bharti:overlaying:2015} and dynamic~\cite{feng:bimodal:2018,deville:dynamic:2014,khodabandelou:metadata:2019,douglass:resolution:2015,liu:dynamics:2018,botta:crowd:2015} population estimation. Power law models are used for mapping mobile phone activity to both static and dynamic population~\cite{bharti:overlaying:2015,feng:bimodal:2018,deville:dynamic:2014,khodabandelou:metadata:2019,douglass:resolution:2015,liu:dynamics:2018}. However, they cannot give consistent population estimations under different spatial and temporal resolutions, as explained in Section~\ref{section:method:rationale}. Nevertheless, human mobility data have demonstrated the great potential for fine dynamic population estimation at an hourly level.


Typically, static population estimation is validated using census data or gridded population datasets~\cite{sutton:ambient:2003,stevens:disaggregating:2015,khodabandelou:metadata:2019,bharti:overlaying:2015,patel:geotweet:2017,yao:building:2017,biljecki:netherlands:2016,steiger:twitter:2015}. It is much more challenging to validate dynamic population estimation due to lack of ground truth. Since dynamic population is found highly correlated with census at night, some researchers validate dynamic population estimation at night using census data or gridded population datasets~\cite{feng:bimodal:2018,deville:dynamic:2014,douglass:resolution:2015,liu:dynamics:2018}. However, this does not justify population dynamics in the whole day. Liu et al. clustered areas based on their similarity in population time series in order to find dynamic population patterns of human mobility~\cite{liu:dynamics:2018}. Khodabandelou et al. visualized dynamic population distributions at different times in a day and explained them with intuition~\cite{khodabandelou:metadata:2019}. Botta et al. correlated dynamic population with the number of attendees to football matches at Stadio San Siro and the number of flights at Linate Airport~\cite{botta:crowd:2015}.


Table~\ref{table:related} summarizes related work in terms of input data sources, validation methods, temporal resolutions and models mapping input to population.
\begin{table*}[htb]
	\centering
	\begin{tabular}{l|l|l|l|l}
		\hline
		\thead{Reference} & \thead{Source} & \thead{Validation} & \thead{Temporal resolution} & \thead{Model} \\
		\hline \hline
		Sutton~\cite{sutton:ambient:2003} & \makecell{ambient, \\ nighttime satellite imagery} & census & static & extrapolation, power law \\
		\hline
		Stevens~\cite{stevens:disaggregating:2015} & \makecell{socioeconomic and \\ morphological datasets} & \makecell{GPW, GRUMP, \\ Afri/AsiaPop} & static & random forest \\
		\hline
		Bakillah~\cite{bakillah:openstreetmap:2014} & \makecell{census, land use, land cover, \\ POIs} & census & static & dasymetric \\
		\hline
		Patel~\cite{patel:geotweet:2017} & geotagged tweets & census & static & random forest \\
		\hline
		Yao~\cite{yao:building:2017} & \makecell{Baidu PoI and \\ Tencent user density} & census & static & \makecell{random forest and gravity} \\
		\hline
		Bharti~\cite{bharti:overlaying:2015} & \makecell{satellite imagery, \\ call data records} & NA & static & power law \\
		\hline
		Biljecki~\cite{biljecki:netherlands:2016} & 3D city models & census & static & multiple linear regression \\
		\hline
		Steiger~\cite{steiger:twitter:2015} & geotagged tweets & census & day and night & NA \\
		\hline
		Deville~\cite{deville:dynamic:2014} & mobile phone data & nighttime against census & season & power law \\
		\hline
		Feng~\cite{feng:bimodal:2018} & mobile phone data & WorldPop & 10, 30 or 60 min & bimodal (power law) \\
		\hline
		Khodabandelou~\cite{khodabandelou:metadata:2019} & mobile network metadata & \makecell{census, \\ dynamic visualization, \\ large events} & 15 min & power law \\
		\hline
		Douglass~\cite{douglass:resolution:2015} & \makecell{telecommunications measures, \\ land use measures} & census & hour & power law, random forest \\
		\hline
		Liu~\cite{liu:dynamics:2018} & call detail records & census, clustering & hour & power law \\
		\hline
		Botta~\cite{botta:crowd:2015} & \makecell{mobile phone activity, \\ Twitter activity} & \makecell{the number of attendees \\ to football matches, \\ the number of flights} & 10 min & ordinary least-squares regression \\
		\hline
	\end{tabular}
	\caption{Summary of related work}\label{table:related}
\end{table*}

This paper focuses on dynamic population estimation using anonymized mobility data. In contrast to~\cite{feng:bimodal:2018,deville:dynamic:2014,khodabandelou:metadata:2019,douglass:resolution:2015,liu:dynamics:2018,botta:crowd:2015} where mobile phone activity data were used, human mobility data collected from a variety of devices are used in this paper. The census data are considered as prior knowledge, while any static population dataset serves the same purpose. A Bayesian model combines static population data and anonymized mobility data for dynamic population estimation. While we demonstrate the results at a spatial resolution of 1 km and a temporal resolution of an hour, it is straightforward to achieve a finer resolution both in space and in time.

This paper is organized as follows. Our method for estimating dynamic population is described in Section~\ref{section:method}. The rationale behind our population model is explained for consistent estimations given different spatial and temporal partitioning schemes. A Bayesian model is proposed based on the rationale for combining static population data and anonymized mobility data. The results of our dynamic population estimation are presented in Section~\ref{section:result}. Finally, the paper is concluded in Section~\ref{section:conclusion}.

\section{Method}\label{section:method}


We use proprietary mobility data from HERE Technologies, an international provider in digital maps and location services. HERE Technologies collects billions of anonymized mobility data from a diversity of devices every day across the globe~\cite{verendel:traffic:2019}. These anonymized mobility data are in the form of short trajectories. A trajectory is a sequence of timestamped GPS coordinates, i.e. probes. The time duration and the sampling frequency may vary from trajectory to trajectory because various data pseudonymization and anonymization technologies are used for privacy protection. While GPS positions are considered in this paper, it is straightforward to extend our methods to the scenarios using different positioning technologies, e.g. indoor positioning.



\subsection{Data Analysis}

\begin{figure}[h]
	\centering
	\includegraphics[width=0.5\columnwidth]{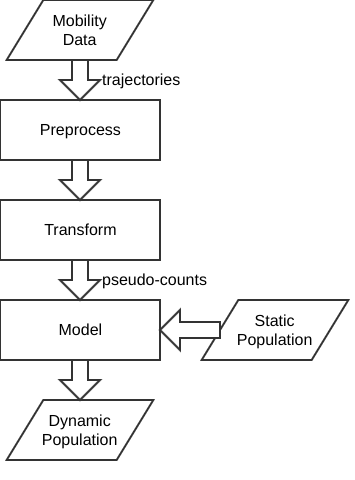}
	\caption{Data Analysis Pipeline for Dynamic Population Estimation}
	\label{figure:pipeline}
\end{figure}

The data analysis pipeline, as shown in Figure~\ref{figure:pipeline}, is used for estimating dynamic population. First, mobility data are preprocessed in order to suppress noise and remove outliers which may be due to cold start, signal drift and loss, etc. Various preprocessing technologies for spatial trajectories have already been investigated in~\cite{lee:preprocessing:2011}. Then, the preprocessed trajectories are transformed into pseudo-counts in each spatiotemporal partition, which is described in Section~\ref{section:method:transform}. The pseudo-count implies the number of observed devices in each spatiotemporal partition. Finally, a model combines static population data and pseudo-counts into dynamic population for all spatiotemporal partitions, which is described in Section~\ref{section:method:model}.

\subsection{Transform}\label{section:method:transform}

A spatial partitioning scheme is denoted as $\mathbf{\Psi}$, which divides a spatial area, e.g. the earth or a city, into a number of disjoint sub-areas. A spatial partitioning scheme may be hierarchical, e.g. HERE tiles~\cite{here:partitions}, Google S2~\cite{google:s2} and Uber H3~\cite{brodsky:h3:2018}, or be defined by a number of polygons, e.g. the district boundaries in Berlin, Germany~\cite{berlin:boundary}. A spatial partition is denoted as $s$, $s \in \mathbb{S}$, where $\mathbb{S}$ is the set of all spatial partitions under the spatial partitioning scheme $\mathbf{\Psi}$ and $|\mathbb{S}|$ is the number of spatial partitions. All spatial partitions under the same spatial partitioning scheme is disjoint, i.e. $s_i \cup s_j = \emptyset$ for $i \neq j$.

A temporal partitioning scheme is denoted as $\mathbf{\Theta}$, which partitions time into intervals having a given length, e.g. 1-hour. A temporal partition is denoted as $t$, $t \in \mathbb{T}$, where $\mathbb{T}$ is the set of all temporal partitions under the temporal partitioning scheme $\mathbf{\Theta}$ and $|t|$ is the length of the temporal partition $t$, e.g. 1 hour.

The preprocessed trajectories are transformed into pseudo-counts in each spatiotemporal partition. The pseudo-count in the spatial partition $s$ and the temporal partition $t$, i.e. in the spatiotemporal partition $(s,t)$, is denoted as $c(s,t)$.

The simplest transform counts the number of probes of all preprocessed trajectories in each spatiotemporal partition. It implies that each probe represents an observed device. This approach has a few problems. Firstly, a device may generate multiple probes within a temporal partition, e.g. 3600 probes in an hour, which substantially over-estimates the number of observed devices. Secondly, the frequency of probe generation varies from device to device. As a consequence, the spatiotemporal partitions where some devices generate probes more frequently would have a disproportionately high amount of pseudo-counts. Thirdly, a device may cross several spatial partitions without generating probes in all of them. For example, a device travels fast on a highway and spatial partitions are relatively small, so that a probe may appear in every second spatial partition.

Another simple transform counts the number of short trajectories in each spatiotemporal partition, which assumes that each trajectory represents an observed device. Apart from the third aforementioned problem, this approach still suffers from two limitations. Firstly, there is no guarantee whether any two trajectories are generated by the same device or not in order to protect privacy as required by GDPR~\cite{eu:gdpr:2016}. The number of short trajectories generated in the same time duration vary from device to device and even from time to time. Consequently, the spatiotemporal partitions, where some devices generate more short trajectories, would have a disproportionately high amount of pseudo-counts. Secondly, a device may traverse a number of spatial partitions in a time partition, which depends on its speed and the spatiotemporal partitioning scheme. Therefore, excessive number of devices are observed in the spatial partitions where many devices travel at high speed, e.g. on highways.

The transform proposed in this paper calculates the intersection between a short trajectory and each spatiotemporal partition, then sums the dwell time of all short trajectories in each spatiotemporal partition. This implies that each observed device is weighted by its dwell time in each spatiotemporal partition, which overcomes the aforementioned problems.

\subsection{Rationale}\label{section:method:rationale}

A model estimates dynamic population from pseudo-counts in each spatiotemporal partition. The population in the spatiotemporal partition $(s,t)$ is denoted as $d(s,t)$.


Power law models are the most popular ones for mapping mobile phone activity to population~\cite{bharti:overlaying:2015,feng:bimodal:2018,deville:dynamic:2014,khodabandelou:metadata:2019,douglass:resolution:2015,liu:dynamics:2018}. They assume a non-linear relationship between the total number of pseudo-counts and the population in each spatiotemporal partition. However, they cannot give consistent estimations when the spatial or temporal partitioning scheme changes, which is explained in detail by Proposition~\ref{proposition:pseudo-count:spatial}-\ref{proposition:population:temporal}.

The rationale behind our model is a few propositions listed below, which are rooted from the intuition of a model for consistent population estimations under different spatial and temporal partitioning schemes.

\begin{proposition}\label{proposition:pseudo-count:spatial}
	Pseudo-counts are additive spatially, i.e.
	\begin{equation}
	c(s_i \cup s_j,t) = c(s_i, t) + c(s_j, t), \quad \forall i \neq j.
	\end{equation}
\end{proposition}
Conceptually, pseudo-count is the number of observed devices inferred from mobility data. The total pseudo-count of two spatial partitions $s_i$ and $s_j$ within the same temporal partition $t$ must be $c(s_i, t) + c(s_j, t)$. Proposition~\ref{proposition:pseudo-count:spatial} suggests that the number of observed devices in a spatial area is the same regardless of the spatial partitioning schemes. For example, no matter if a city is partitioned into 100 or 200 sub-areas, its pseudo-count remains the same, which is simply the sum of the pseudo-counts in all sub-areas.

\begin{proposition}\label{proposition:pseudo-count:temporal}
	Pseudo-counts are additive temporally, i.e.
	\begin{equation}
		c(s,t_i \cup t_j) = c(s, t_i) + c(s, t_j), \quad \forall i \neq j.
	\end{equation}
\end{proposition}
Similarly, the total pseudo-count of two temporal partitions $t_i$ and $t_j$ in the same spatial partition $s$ must be $c(s, t_i) + c(s, t_j)$. Proposition~\ref{proposition:pseudo-count:temporal} suggests that the number of observed devices during a time interval is the same regardless of the temporal partitioning schemes. Proposition~\ref{proposition:pseudo-count:temporal} also implies that each observed device should be weighted by its dwell time in each spatial partition, otherwise a device appearing within the temporal partitions $t_i$ and $t_j$ would be counted as two devices within the temporal partition $t_i \cup t_j$.

\begin{proposition}\label{proposition:population:spatial}
	Populations are additive spatially, i.e.
	\begin{equation}
	d(s_i \cup s_j, t) = d(s_i, t) + d(s_j, t), \quad \forall i \neq j.
	\end{equation}
\end{proposition}
Similar to Proposition~\ref{proposition:pseudo-count:spatial}, Proposition~\ref{proposition:population:spatial} suggests that the population in a spatial area is the same regardless of the spatial partitioning schemes.

\begin{proposition}\label{proposition:population:temporal}
	The population within two temporal partitions is between the populations of these two individual temporal partitions, i.e.
	\begin{align}\label{equation:population:temporal}
	\min\{d(s, t_i), d(s, t_j)\} & \leq d(s, t_i \cup t_j) \nonumber \\
	& \leq \max\{d(s, t_i), d(s, t_j)\}, \forall i \neq j.
	\end{align}
\end{proposition}
Proposition~\ref{proposition:population:temporal} suggests that the population $d(s, t_i \cup t_j)$ is a ``middle'' value between the populations $d(s, t_i)$ and $d(s, t_j)$. It turns out that mediant~\cite{guhery:mediant:2010} is used in our model, which is explained in Section~\ref{section:method:model}.

\begin{proposition}\label{proposition:prior}
	When the pseudo-counts of some spatial partitions are small, i.e. few devices are observed, within a temporal partition, their populations within the temporal partition needs some prior knowledge.
\end{proposition}
Proposition~\ref{proposition:prior} suggests a Bayesian model. The prior knowledge used in this paper is census data, but it could be any static population datasets. It makes sense that the best guess is the static population when there are no mobility data at all.

\begin{proposition}\label{proposition:likelihood}
	When the pseudo-counts of some spatial partitions are large, i.e. many devices are observed, within a temporal partition, their populations within the temporal partition are roughly proportional to their pseudo-counts, i.e.
	\begin{equation}
	\frac{d(s_i,t)}{c(s_i,t)} \approx \frac{d(s_j,t)}{c(s_j,t)}, \quad \forall i \neq j.
	\end{equation}
\end{proposition}
Proposition~\ref{proposition:likelihood} suggests that mobility data more or less reflects population distribution in the daytime. But it does not apply to pseudo-counts within different temporal partitions, i.e. typically
\begin{equation}
\frac{d(s,t_i)}{c(s,t_i)} \not\approx \frac{d(s,t_j)}{c(s,t_j)} , \quad \forall i \neq j.
\end{equation}
For example, there are much less devices observed in entire Germany in the nighttime than in the daytime, while the total population may stay almost the same.

\subsection{Model}\label{section:method:model}

Based on the rationale in Section~\ref{section:method:rationale}, a Bayesian model is proposed in this paper for estimating the probability observing a device in all spatial partitions within a given temporal partition.

The likelihood function within the given temporal partition $t$ is a categorical distribution. The categories are $|\mathbb{S}|$ spatial partitions under a spatial partitioning scheme $\mathbf{\Psi}$, hence ``category'' and ``spatial partition'' are interchangeable hereafter in this paper. The probability of the category $s$ within the temporal partition $t$ represents the probability that a device is observed in the spatial partition $s$.

The conjugate prior distribution within the given temporal partition $t$ is a Dirichlet distribution having the concentration parameters $\alpha(s,t) \geq 0$ for $s = 1,\dots,|\mathbb{S}|$. Intuitively, the prior concentration parameters $\alpha(s,t)$ are pseudo-counts from the prior knowledge, which are derived from the census data in this paper. Typically, the census data, e.g. in Germany~\cite{germany:census:2011}, use a spatial partitioning scheme different from the desired one, i.e. $\mathbf{\Psi}$. Therefore, the census population has to be firstly disaggregated to all spatial partitions under $\mathbf{\Psi}$. A variety of disaggregation methods have been comprehensively reviewed in~\cite{wu:population:2005}. Such a population disaggregation from one spatial partitioning scheme to another implicitly assumes Proposition~\ref{proposition:population:spatial}. After disaggregation, the population in the spatial partition $s$ is denoted as $b(s)$. Although the total population of all spatial partitions in a considered area varies over the time in reality, the fluctuation is ignored compared to the total population if the considered area is large enough, e.g. entire Germany. The prior concentration parameters $\alpha(s,t)$ can be calculated by
\begin{equation}\label{equation:prior:concentration}
\alpha(s,t) = \lambda b(s) |t|,
\end{equation}
where $\lambda$ is a scaling factor which balances the prior distribution and the likelihood function.

The posterior distribution is also Dirichlet, which has the posterior concentration parameters
\begin{equation}\label{equation:posterior:concentration}
\hat{\alpha}(s,t) = \alpha(s,t) + c(s,t).
\end{equation}
As a result, the posterior mean is
\begin{equation}\label{equation:posterior:mean}
\mathbb{E}\{\hat{p}(s,t)\} = \frac{\hat{\alpha}(s,t)}{\displaystyle \sum_{s'=1}^{|\mathbb{S}|} \hat{\alpha}(s',t)},
\end{equation}
where $\hat{p}(s,t)$ is the posterior probability observing a device in the spatiotemporal partition $(s,t)$.

Finally, the estimated population $\hat{d}(s,t)$ can be computed as
\begin{equation}
\hat{d}(s,t) = N \mathbb{E}\{\hat{p}(s,t)\},
\end{equation}
where $N$ is the total population of all spatial partitions in the considered area,
\begin{equation}\label{equation:population:estimation}
N = \sum_{s=1}^{|\mathbb{S}|} b(s).
\end{equation}

The proof that our model meets Proposition~\ref{proposition:population:temporal} is as follows, while the other propositions in Section~\ref{section:method:rationale} are straightforward.
\begin{proof}
	The estimated population in the spatiotemporal partition $(s,t_i \cup t_j)$, i.e. $\hat{d}(s, t_i \cup t_j)$, can be derived from Equations~\ref{equation:prior:concentration}, \ref{equation:posterior:concentration}, \ref{equation:posterior:mean} and \ref{equation:population:estimation}:
	\begin{align}
	\hat{d}(s, t_i \cup t_j) = N \frac{
		\lambda b(s) \left( |t_i| + |t_j| \right) + \left[ c(s, t_i) + c(s, t_j) \right]
	}{
		\displaystyle \lambda N \left( |t_i| + |t_j| \right) + \sum_{s'=1}^{|\mathbb{S}|} \left[ c(s', t_i) + c(s', t_j) \right]
	},
	\end{align}
	which is the mediant of $d(s, t_i)$ and $d(s, t_j)$. Inequality~\ref{equation:population:temporal} is satisfied according the mediant inequality~\cite{guhery:mediant:2010}.
\end{proof}

\section{Result}\label{section:result}

\begin{figure}[h]
	\centering
	\subfloat[Census\label{figure:density:census}]{%
		\includegraphics[width=0.49\linewidth]{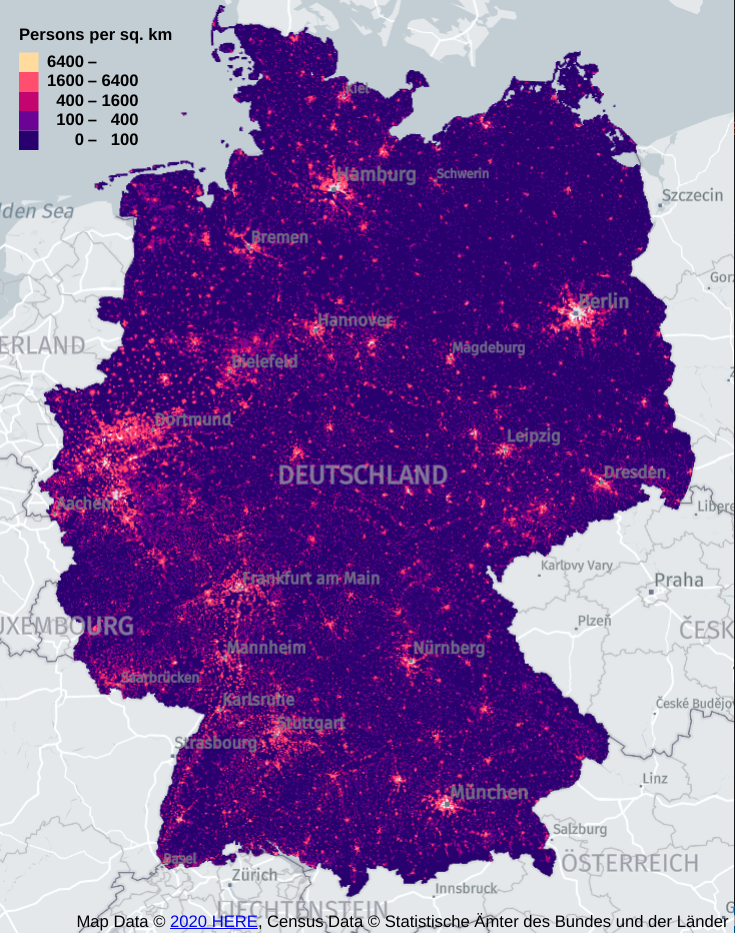}}
	\hfill
	\subfloat[Pseudo-Count\label{figure:density:probe}]{%
		\includegraphics[width=0.49\linewidth]{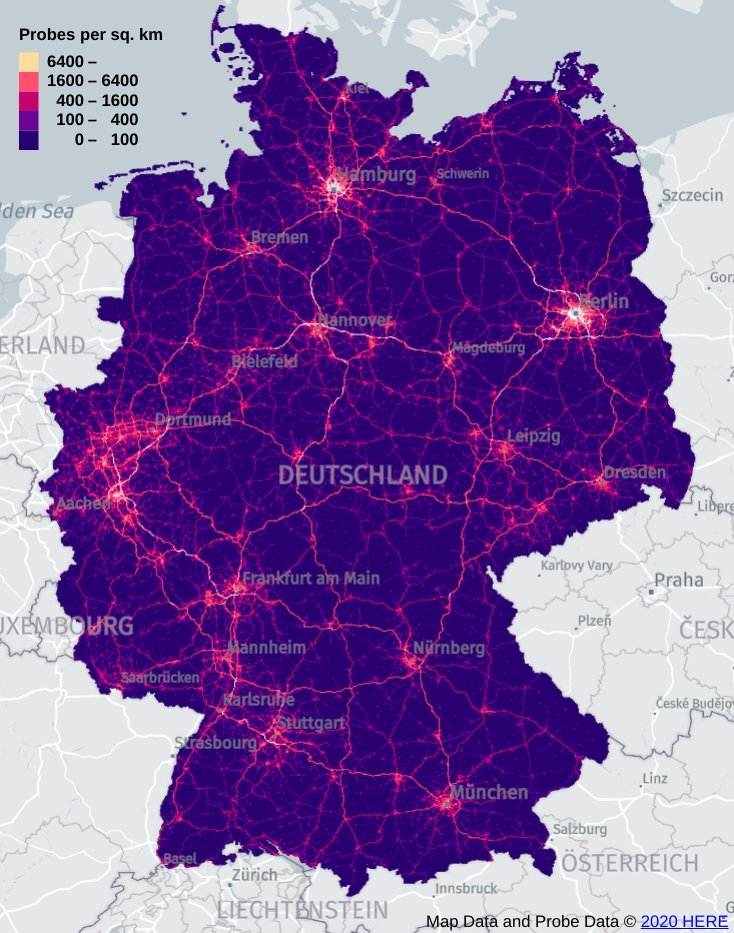}}
	\caption{The Census Population Density and Pseudo-Count Density in Germany}
	\label{figure:density}
\end{figure}

The Germany census data with 1~km resolution~\cite{germany:census:2011} is used in this paper as the prior knowledge, where the entire Germany is partitioned into 361,478 squares. Figure~\ref{figure:density:census} shows the spatial distribution of census population density in Germany.

The anonymized mobility data is provided by HERE Technologies. There are more than 8.4 billion probes, which covers the whole Germany from 1 April to 7 April 2019 in UTC.

Figure~\ref{figure:density:probe} shows the spatial distribution of pseudo-count density in Germany. Pseudo-counts are also partitioned into the same squares as those used in the census data in order to avoid disaggregation. The timestamps are ignored. The total number of pseudo-counts is scaled to the total population for the sake of comparison with census population density.

Big cities are clearly visible in both Figure~\ref{figure:density:census} and~\ref{figure:density:probe}, because more human activities are observed in densely populated territories. Additionally, Figure~\ref{figure:density:probe} also identifies highways and arterial roads due to large amount of human activities observed there. This confirms that mobility data provides a very important complement for dynamic population estimation.

\begin{figure}[h]
	\centering
	\subfloat[Census\label{figure:density:census:berlin}]{%
		\includegraphics[width=0.49\linewidth]{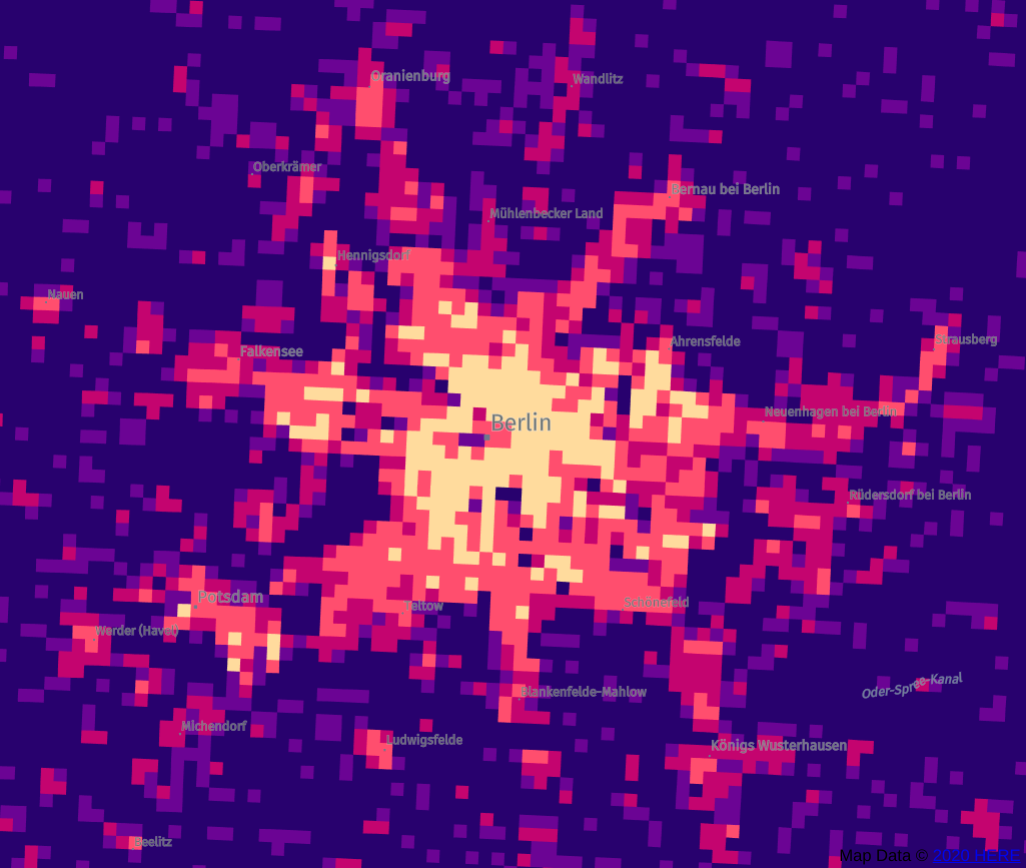}}
	\hfill
	\subfloat[Pseudo-Count\label{figure:density:probe:berlin}]{%
		\includegraphics[width=0.49\linewidth]{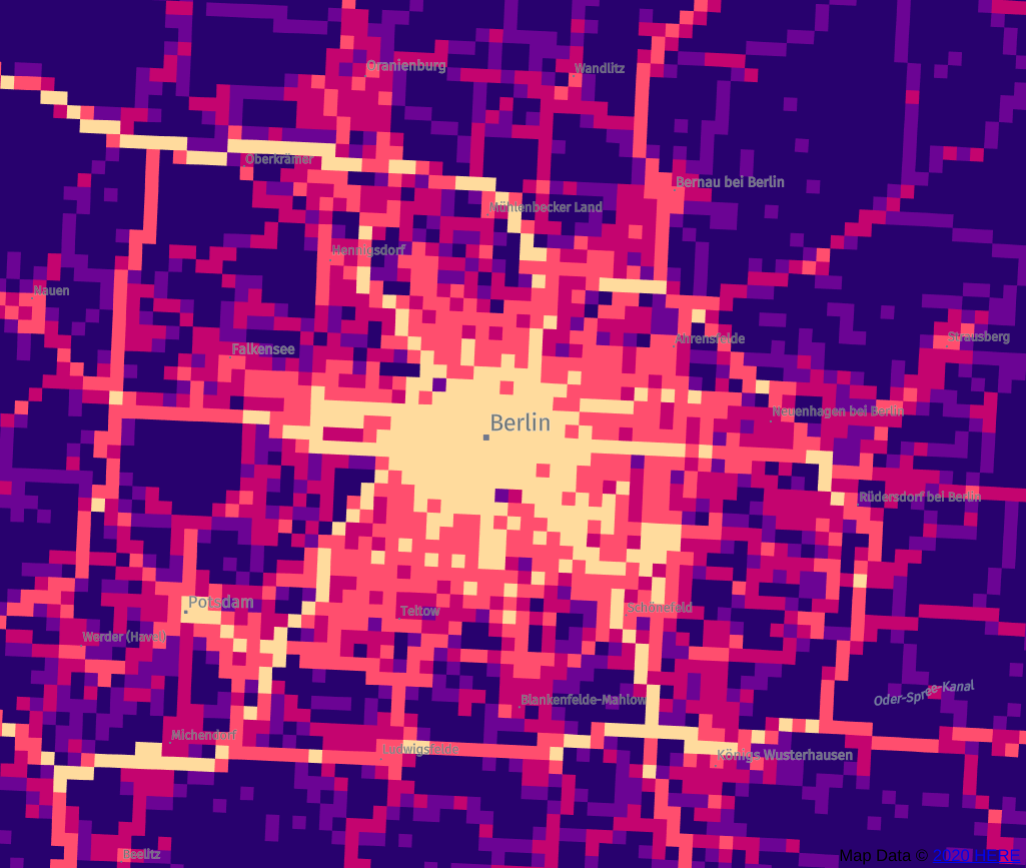}}
	\caption{The Census Population Density and Pseudo-Count Density in Berlin}
	\label{figure:density:berlin}
\end{figure}

Figure~\ref{figure:density:census:berlin} and~\ref{figure:density:probe:berlin} show the spatial distribution of census population density and pseudo-count density around Berlin by zooming in on Figure~\ref{figure:density:census} and~\ref{figure:density:probe}, respectively. It is evident that the orbital highway A10 around Berlin has a high pseud-count density in yellow.

\begin{figure}[h]
	\centering
	\includegraphics[width=\linewidth]{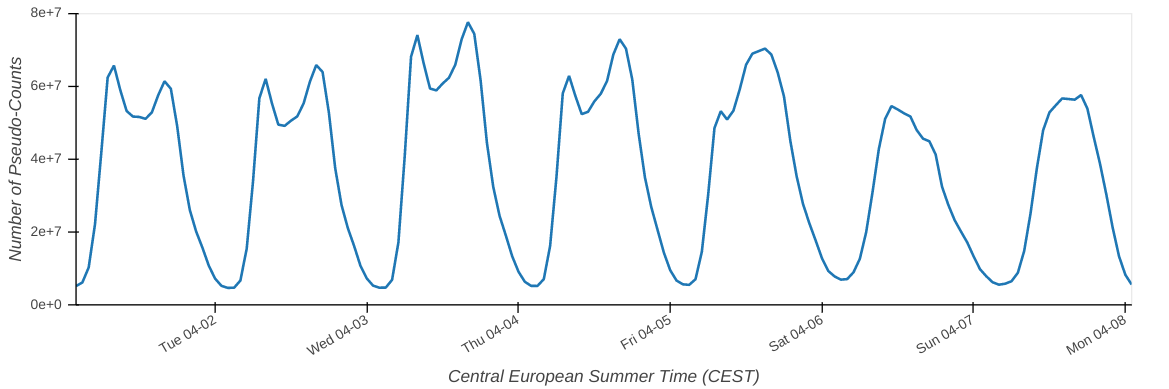}
	\caption{The Total Number of Pseudo-Counts within Each Hour in Germany.}
	\label{figure:time:probe}
\end{figure}

Figure~\ref{figure:time:probe} shows the total number of pseudo-counts within each hour in Germany, It shows a rush hour pattern around 8:00 and 16:00 CEST during weekdays, which is consistent with human activities. The number of pseudo-counts between 16:00 and 17:00 on 3 April 2019 is 15 times as much as that between 2:00 and 3:00 on 2 April 2019. However, it is impossible that the population in Germany from 1 April to 7 April 2019 could fluctuate at the same magnitude. Hence, a simple linear model such as $y_t=a x_t$ with a constant scaling factor $a$ for all temporal partitions is inappropriate. 

\begin{figure}[h]
	\centering
	\includegraphics[width=\linewidth]{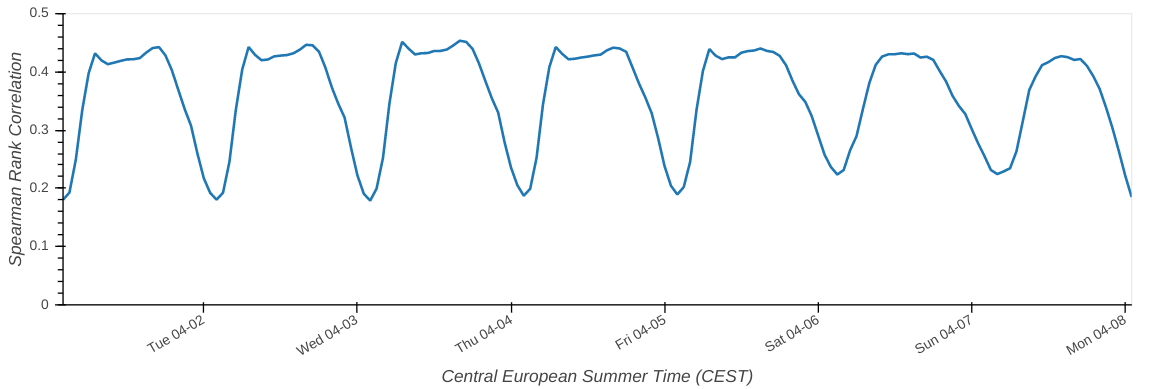}
	\caption{Spearman Rank Correlation between the Number of Pseudo-Counts and the Census Population within Each Hour in Germany.}
	\label{figure:spearman}
\end{figure}

Figure~\ref{figure:spearman} shows Spearman rank correlation between the number of pseudo-counts and the census population within each hour in Germany. In contrast to mobile phone data which are highly correlated with census population at night~\cite{deville:dynamic:2014,khodabandelou:metadata:2019,douglass:resolution:2015,liu:dynamics:2018,botta:crowd:2015}, mobility data are only weakly correlated with census population while the correlation is higher during the daytime than that during the nighttime. Therefore, mobility data alone cannot be used for dynamic population estimation.

\begin{figure}[h]
	\centering
	\subfloat[02:00-03:00\label{figure:estimation:spatial:02h}]{%
		\includegraphics[width=0.49\linewidth]{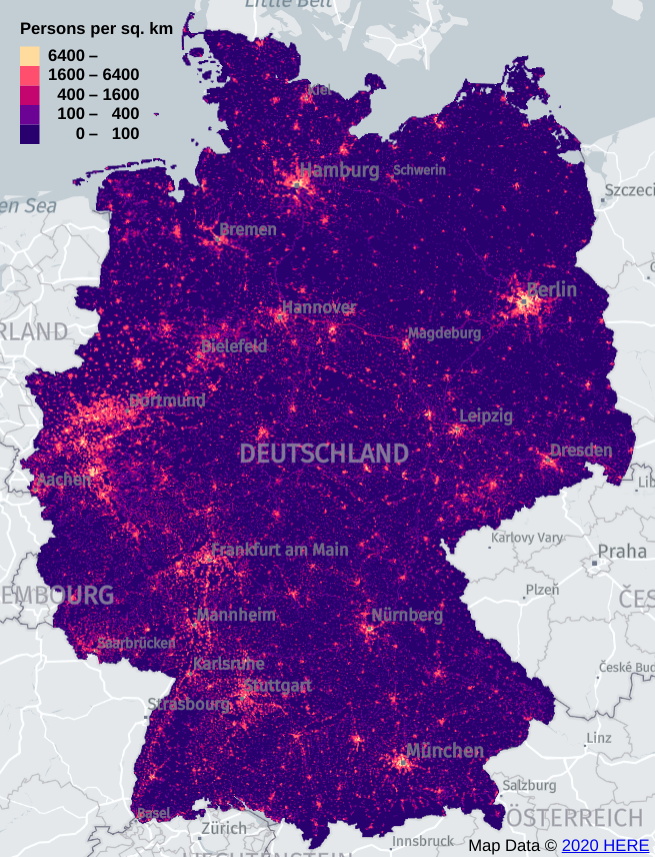}}
	\hfill
	\subfloat[14:00-15:00\label{figure:estimation:spatial:14h}]{%
		\includegraphics[width=0.49\linewidth]{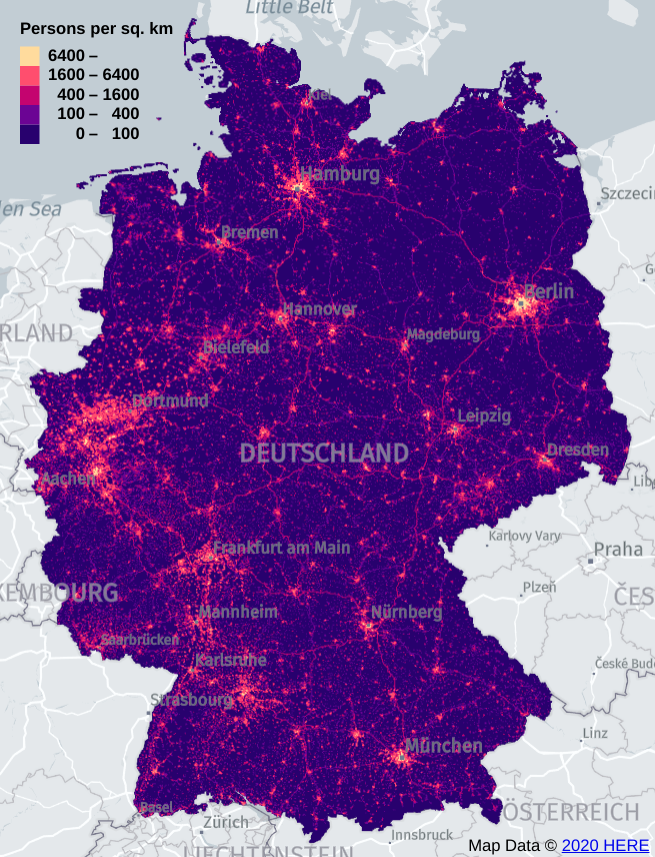}}
	\caption{The Estimated Population Density in Germany on 1 April 2019}
	\label{figure:estimation:spatial}
\end{figure}

\begin{figure}[h]
	\centering
	\subfloat[02:00-03:00\label{figure:estimation:spatial:02h:berlin}]{%
		\includegraphics[width=0.49\linewidth]{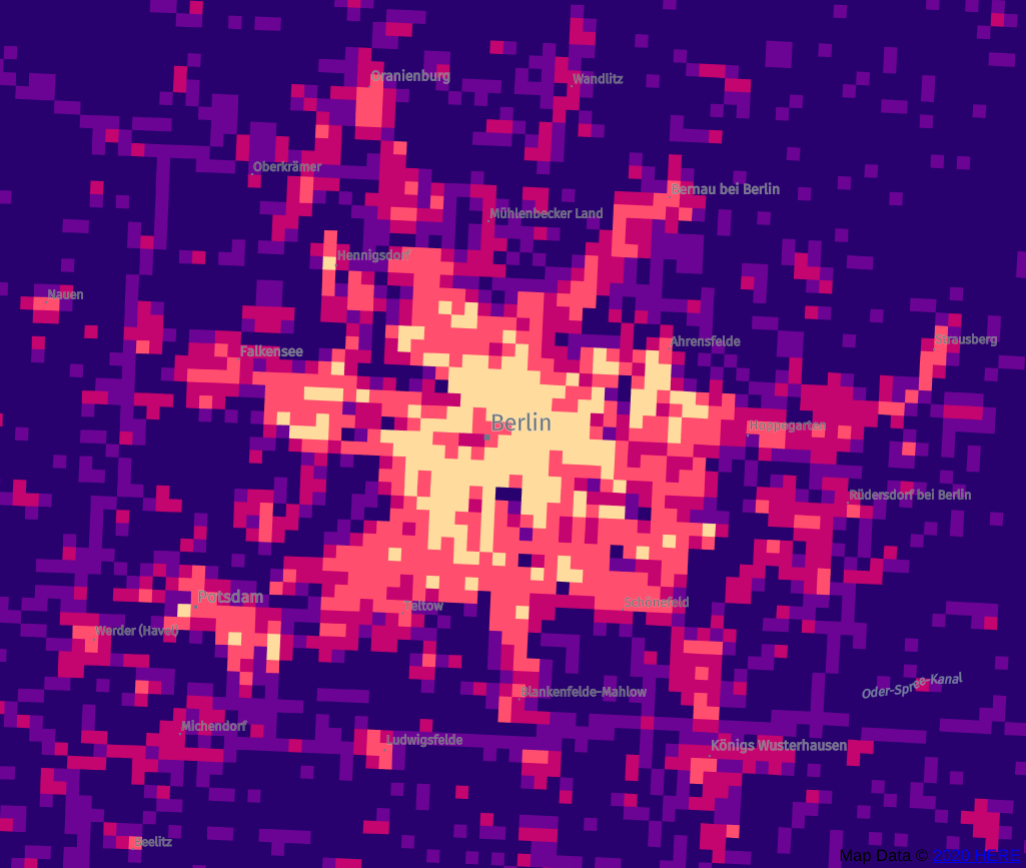}}
	\hfill
	\subfloat[14:00-15:00\label{figure:estimation:spatial:14h:berlin}]{%
		\includegraphics[width=0.49\linewidth]{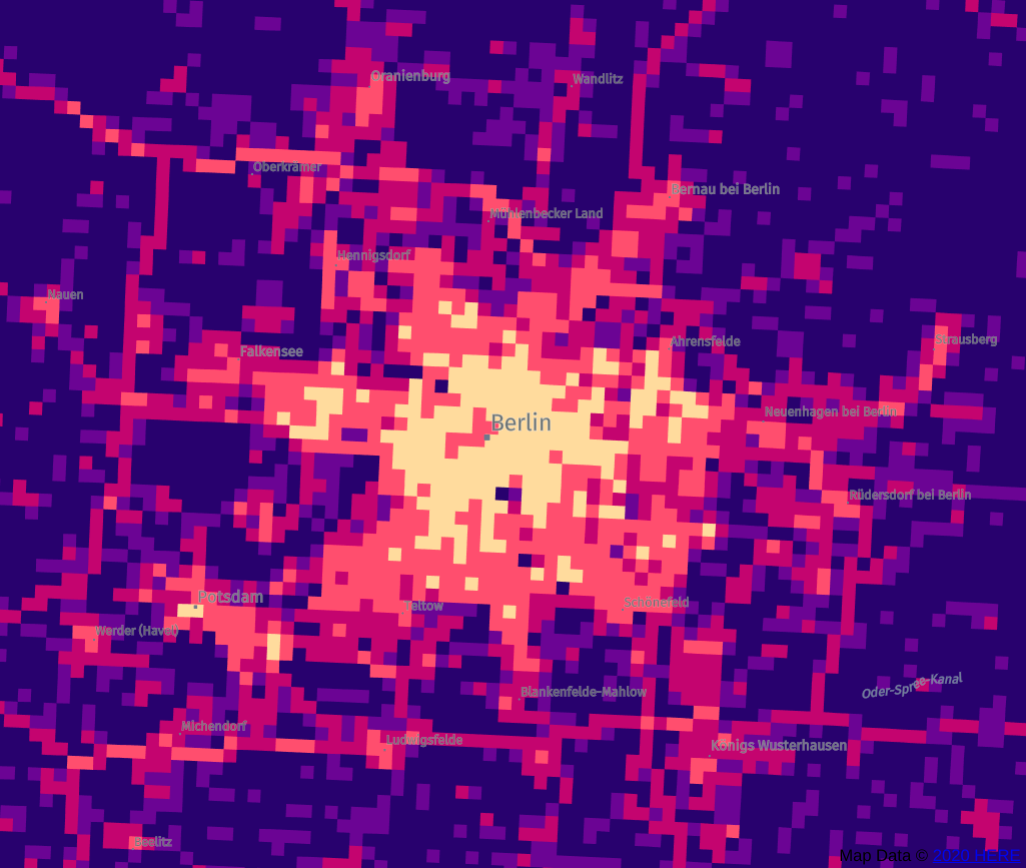}}
	\caption{The Estimated Population Density in Berlin on 1 April 2019}
	\label{figure:estimation:spatial:berlin}
\end{figure}

Figure~\ref{figure:estimation:spatial} shows the spatial distribution of our estimated population density in Germany on 1 April 2019 in the hours starting from 2:00 and 14:00, respectively. Figure~\ref{figure:estimation:spatial:berlin} shows the same spatial distribution of our estimated population density in Berlin by zooming in on Figure~\ref{figure:estimation:spatial}. The estimated nighttime population density in Figure~\ref{figure:estimation:spatial:02h} and~\ref{figure:estimation:spatial:02h:berlin} is similar to the census population density in Figure~\ref{figure:density:census} and~\ref{figure:density:census:berlin}, because there are not much mobility data at night. The estimated daytime population density in Figure~\ref{figure:estimation:spatial:14h} and~\ref{figure:estimation:spatial:14h:berlin} is improved compared to the pseud-count density in Figure~\ref{figure:density:probe} and~\ref{figure:density:probe:berlin}. The highways and arterial roads are still identified, but with lower estimated population density. Compared to Figure~\ref{figure:density:probe:berlin}, Figure~\ref{figure:estimation:spatial:14h:berlin} shows a much lower population density in red or orange on the orbital highway A10 around Berlin in the afternoon.

The dynamic population estimation of each square from 1 April to 7 April 2019 is a time series, which is further normalized to z-scores. All squares are clustered based on the similarity between these z-score time series using hdbscan~\cite{mcinnes:accelerated:2017,mcinnes:hdbscan:2017} with the metric $\sqrt{P}$earson dissimilarity~\cite{solo:pearson:2019,gower:dissimilarity:1986}. The $\sqrt{P}$earson dissimilarity between the population time series of two spatial partitions, i.e. $s_i$ and $s_j$, is defined as $\sqrt{1-\rho_{i,j}}$, where $\rho_{i,j}$ is the Pearson correlation between the population time series.

\begin{figure}[h]
	\centering
	\includegraphics[width=0.7\linewidth]{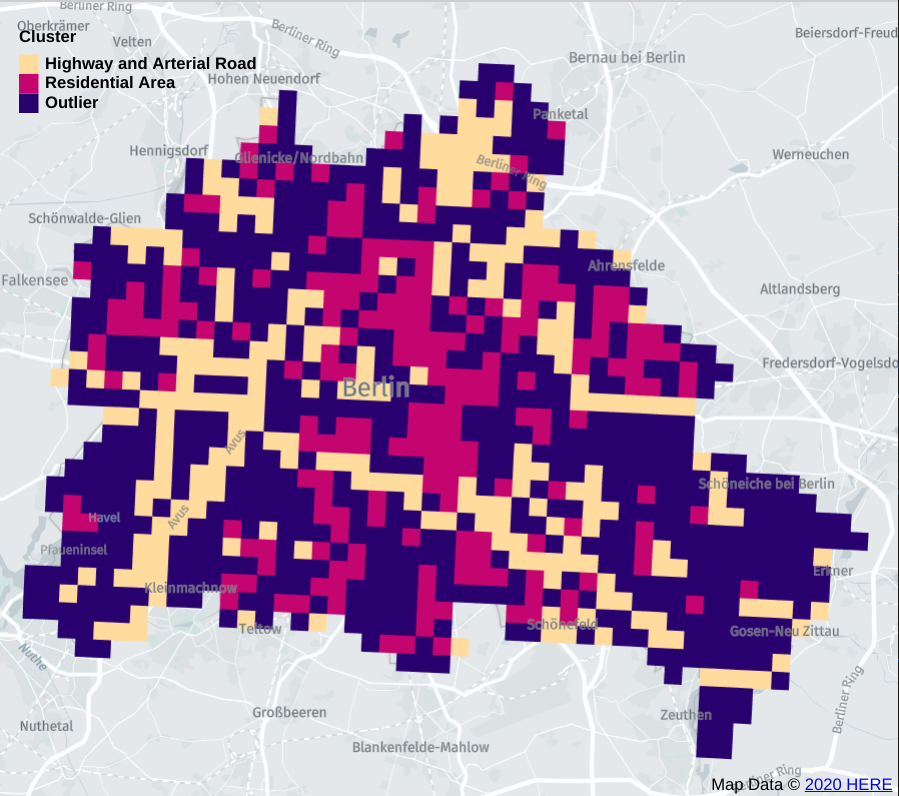}
	\caption{The Spatial Distribution of Estimated Dynamic Population Patterns in Berlin}
	\label{figure:berlin:cluster}
\end{figure}

Figure~\ref{figure:berlin:cluster} shows the spatial distribution of the clusters in Berlin. Most parts of the highways (e.g. A10, A100, A111, A113, A114 and A115) and the arterial roads (e.g. B1, B2 and B96a) are identified. Many residential areas in Prenzlauer Berg, Kreuzberg and Neukölln are also identified. In addition, there are many outliers, which might be due to large squares mixing different functional areas or simply insufficient mobility data in one week.

\begin{figure}[h]
	\centering
	\includegraphics[width=\linewidth]{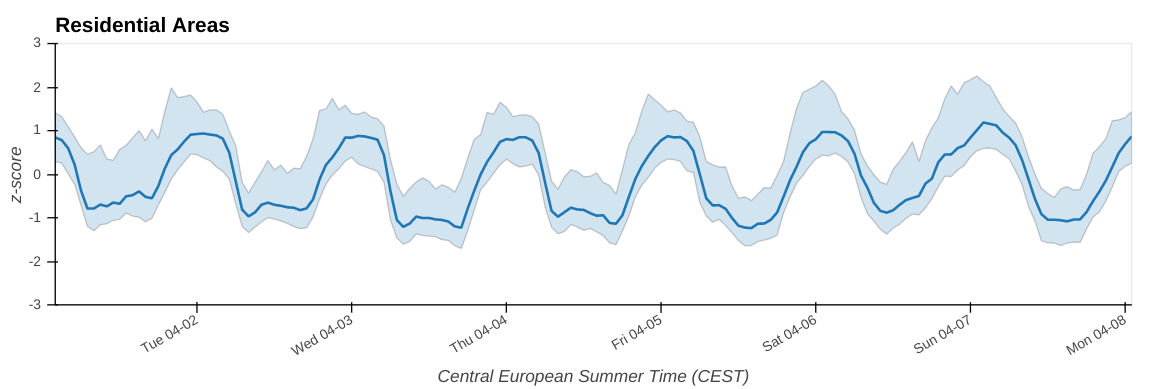}
	\\
	\includegraphics[width=\linewidth]{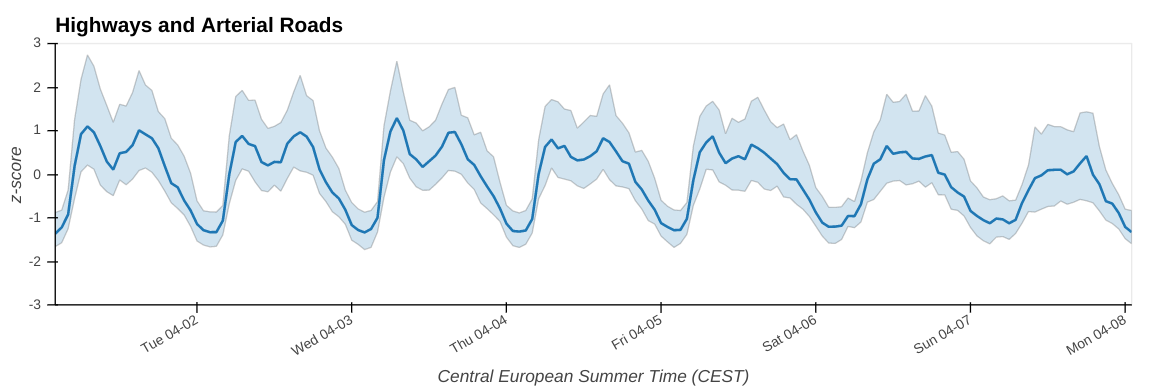}
	\caption{The z-score Time Series}
	\label{figure:berlin:zscore}
\end{figure}

Figure~\ref{figure:berlin:zscore} shows the z-score time series of the two major clusters, where the line in the middle represents the median z-score while the shadow represents the 10th and 90th percentiles. The residential pattern is characterised by more people in the nighttime than in the daytime, as suggested by the z-score values. Its population in weekdays decreases between 4:00 and 8:00 and increases between 17:00 and 23:00. The highway and arterial road pattern is characterised by two rush hour peaks in weekdays. Usually its rush hour peaks in weekdays are at 7:00-8:00 and 16:00-17:00, respectively, with a bit earlier peak in the Friday afternoon. Its population in weekend remains high during the daytime without clear rush hour peaks.

\section{Conclusion}\label{section:conclusion}

In this paper a Bayesian model is proposed for dynamic population estimation using anonymized mobility data. The rationale behind our model is consistent population estimations under different spatial and temporal partitioning schemes. The static population data, e.g. census data, is considered as prior knowledge. Our Bayesian model combines the static population data and the anonymized mobility data. Our results at a spatial resolution of 1~km and a temporal resolution of 1~hour are demonstrated. The spatial distributions of the estimated population during the daytime and during the nighttime are visualized, which are consistent with intuition. Two dynamic population patterns are identified from our results, which clearly reveals population dynamics during a week. The results in this paper help us better understand population dynamics at a fine level of spatial and temporal resolutions.

\bibliographystyle{IEEEtran}
\bibliography{IEEEabrv,references}
\end{document}